\documentclass[aps,prb,showpacs, twocolumn,superscriptaddress]{revtex4-2}
\usepackage{bm,color,amsmath,amssymb,mathrsfs,latexsym,graphicx,psfrag,mathtools}
\usepackage{color}
\usepackage[dvipsnames]{xcolor}
\usepackage[hidelinks,colorlinks,linkcolor=blue,
citecolor=blue,urlcolor=blue]{hyperref}
\usepackage{dsfont}

\usepackage{empheq}

\newcommand{\bsl}[1]{\boldsymbol{#1}}

\newcommand{\figref}[1]{Fig.\,\ref{#1}}

\newcommand{\refcite}[1]{Ref.\,\cite{#1}}

\let\oldAA\AA
\renewcommand{\AA}{\text{\normalfont\oldAA}}

\newcommand{\ie}{{\emph{i.e.}}}
\newcommand{\eg}{{\emph{e.g.}}}

\newcommand{\abi}{\text{\emph{ab initio}}}
\newcommand{\pcomm}{Prof.\,Pickett's comment}

\newcommand{\mgb}{\text{MgB$_2$}}

\usepackage{comment}

\begin{document}
\title{Reply to ``Comment on \emph{Nontrivial Quantum Geometry and the Strength of Electron-Phonon Coupling, arXiv:2305.02340}, J. Yu, C. J. Ciccarino, R. Bianco, I. Errea, P. Narang, B. A. Bernevig"}

\author{Jiabin Yu}
\affiliation{Department of Physics, Princeton University, Princeton, NJ 08544, USA}

\author{Christopher J. Ciccarino}
\affiliation{Department of Materials Science and Engineering, Stanford University, CA 94305, USA}
\affiliation{College of Letters and Science, University of California, Los Angeles, CA 90095, USA}

\author{Raffaello Bianco}
\affiliation{Centro de F\'isica de Materiales (CSIC-UPV/EHU), Manuel de Lardizabal pasealekua 5, 20018 Donostia/San Sebasti\'an, Spain}
\affiliation{ Ruder Boskovi\'c Institute, 10000 Zagreb, Croatia}
\affiliation{ Dipartimento di Scienze Fisiche, Informatiche e Matematiche, Universit\`a di Modena e Reggio Emilia, Via Campi 213/a I-41125 Modena, Italy}
\affiliation{ Centro S3, Istituto Nanoscienze-CNR, Via Campi 213/a, I-41125 Modena, Italy}

\author{Ion Errea}
\affiliation{Centro de F\'isica de Materiales (CSIC-UPV/EHU), Manuel de Lardizabal pasealekua 5, 20018 Donostia/San Sebasti\'an, Spain}
\affiliation{Fisika Aplikatua Saila, Gipuzkoako Ingeniaritza Eskola, University of the Basque Country (UPV/EHU), Europa Plaza 1, 20018 Donostia/San Sebasti\'an, Spain}
\affiliation{Donostia International Physics Center (DIPC), Manuel Lardizabal pasealekua 4, 20018 Donostia/San Sebasti\'an, Spain}

\author{Prineha Narang}
\affiliation{College of Letters and Science, University of California, Los Angeles, CA 90095, USA}

\author{B. Andrei Bernevig}
\email{bernevig@princeton.edu}
\affiliation{Department of Physics, Princeton University, Princeton, NJ 08544, USA}
\affiliation{Donostia International Physics Center (DIPC), Manuel Lardizabal pasealekua 4, 20018 Donostia/San Sebasti\'an, Spain}
\affiliation{IKERBASQUE, Basque Foundation for Science, Bilbao, Spain}

\begin{abstract}
    This is our reply to ``Comment on \emph{Nontrivial Quantum Geometry and the Strength of Electron-Phonon Coupling, arXiv:2305.02340}, J. Yu, C. J. Ciccarino, R. Bianco, I. Errea, P. Narang, B. A. Bernevig" by Prof.\,Pickett~\cite{pickett2023comment} (referred to as {\pcomm} in the main text of the reply), which focuses on the {\mgb} part of our work~\cite{Yu05032023GeometryEPC}.
    We show that the entirety of the criticism in \refcite{pickett2023comment} pertaining to our work~\cite{Yu05032023GeometryEPC} is invalid. 
\end{abstract}

\maketitle

We emphasize that the fundamental/basic physics of the electron-phonon coupling (EPC) of {\mgb} have been fully included in our work~\cite{Yu05032023GeometryEPC}. 
For example, in contrast to the claim in the summary paragraph of {\pcomm}, our results derived from our EPC model explicitly demonstrate the colossal impact of the cylindrical Fermi surfaces around $\Gamma$-A (given by $\sigma$-bonding electron states), and the uniquely large EPC matrix elements and scattering processes (from the $E_{2g}$ bond-stretching phonon modes) on/across those Fermi surfaces.
The detailed responses to this point and all other points in {\pcomm} are presented in the appendices.
Our results on ${\mgb}$ are consistent with the previous results (especially from {\abi} calculations) in the literature.
The consistency is evidenced by the good match between the EPC constant from our model and that from the {\abi} calculation for {\mgb} (see TABLE I of our work~\cite{Yu05032023GeometryEPC}), and is also checked in App.\,H6 of our work~\cite{Yu05032023GeometryEPC}.

Our work~\cite{Yu05032023GeometryEPC} provides a new way (from quantum geometry) to understand the EPC.
Quantum geometry describes the real-space localization properties of the electrons (\ie, the Wannier spread), and thus should, on general grounds, affect the strength of the EPC projected to the bands of interests.
The geometric part of EPC (or more specifically EPC constant $\lambda$) is included, but very much hidden, in previous EPC results given by {\abi} calculations (including those on {\mgb}).
Our work provides a way to analytically identify the geometric part (and distinguish it from the energetic part) of the EPC in graphene and {\mgb} (as well as systems/models similar to them).
The approaches, analytical expressions, and numerical results in our work do not contradict any of the known literature in {\mgb}.
Our separation of the geometric and energetic parts of EPC provides a new understanding of the EPC, which further provides a new insight that could help the search for new materials with stronger EPC---\eg, if two systems have similar energetic properties, the one with stronger geometric properties would tend to have stronger EPC.

Uncovering the effects of quantum geometry on the physics of multi-band systems~\cite{Provost1980FSMetric,Resta2011QuantumGeometry} has become a key area of interest in physics. The realization that quantum geometry probably holds an equally fundamental role to that of Berry phases has spurred a large body of work on its crucial effect in various phenomena in flat band systems, including superfluid weight~\cite{Torma2015SWBoundChern,Torma2016SuperfluidWeightLieb,Torma2018SelectiveQuantumMetric,Xie2020TopologyBoundSCTBG,Herzogarbeitman2021SWBound,Verma2021FlatBandSC,Rossi2019SFWTBG,Torma2020SFWTBG,Park2020SCHofBut,Herzog2022FlatBandQuantumGeometry,Huhtinen2022FlatBandSCQuantumMetric, Yu2022EOCPTBG,Herzog2022ManyBodySCFlatBand,Huang2022arXivQuantumGeometryPDW,Law2023QuantumMetricLandau,Torma2022ReviewQuantumGeometry,Chowdhury2022SFFlatBandQuantumGeometry}, the fractional Chern insulators~\cite{BAB2011FCI,Sondhi2013FCI,Regnault2013FCI,Roy2014FCI,Vishwanath2020FCITBG,Wang2022FCITwistedGraphene,Vishwanath2020FCITBG,Wang2022FCITwistedGraphene}, and other phenomena~\cite{Montambaux2016GeometryObitalSusceptibility,Yang2020QuantumDistanceFlatBands,Mera2021FlatBandsKahler,Torma2021FlatBandBEC,Wang2021GeometryFlatBand,Rossi2022QuantumMetricExciton,Holder2022FlatBandQuantumMetricResistivity,Refael2022ShiftCurrentTBG,Oh04252022RhimQuantumDistance}. Our work shows that the quantum geometry of dispersive systems (on Fermi surfaces) is crucial to a more in-depth --- albeit not contradictory to prior work --- understanding of the EPC (strength),  and more generally demonstrates the principle that quantum geometry can affect the strength of realistic interactions in solids.

In the remainder of the reply, we address the points raised by {\pcomm} one by one; all of the points are on {\mgb}, and the entirety of the criticism of our work~\cite{Yu05032023GeometryEPC} is invalid.

\bibliography{bibfile_references.bib}

\begin{thebibliography}{38}%
\makeatletter
\providecommand \@ifxundefined [1]{%
 \@ifx{#1\undefined}
}%
\providecommand \@ifnum [1]{%
 \ifnum #1\expandafter \@firstoftwo
 \else \expandafter \@secondoftwo
 \fi
}%
\providecommand \@ifx [1]{%
 \ifx #1\expandafter \@firstoftwo
 \else \expandafter \@secondoftwo
 \fi
}%
\providecommand \natexlab [1]{#1}%
\providecommand \enquote  [1]{``#1''}%
\providecommand \bibnamefont  [1]{#1}%
\providecommand \bibfnamefont [1]{#1}%
\providecommand \citenamefont [1]{#1}%
\providecommand \href@noop [0]{\@secondoftwo}%
\providecommand \href [0]{\begingroup \@sanitize@url \@href}%
\providecommand \@href[1]{\@@startlink{#1}\@@href}%
\providecommand \@@href[1]{\endgroup#1\@@endlink}%
\providecommand \@sanitize@url [0]{\catcode `\\12\catcode `\$12\catcode
  `\&12\catcode `\#12\catcode `\^12\catcode `\_12\catcode `\%12\relax}%
\providecommand \@@startlink[1]{}%
\providecommand \@@endlink[0]{}%
\providecommand \url  [0]{\begingroup\@sanitize@url \@url }%
\providecommand \@url [1]{\endgroup\@href {#1}{\urlprefix }}%
\providecommand \urlprefix  [0]{URL }%
\providecommand \Eprint [0]{\href }%
\providecommand \doibase [0]{https://doi.org/}%
\providecommand \selectlanguage [0]{\@gobble}%
\providecommand \bibinfo  [0]{\@secondoftwo}%
\providecommand \bibfield  [0]{\@secondoftwo}%
\providecommand \translation [1]{[#1]}%
\providecommand \BibitemOpen [0]{}%
\providecommand \bibitemStop [0]{}%
\providecommand \bibitemNoStop [0]{.\EOS\space}%
\providecommand \EOS [0]{\spacefactor3000\relax}%
\providecommand \BibitemShut  [1]{\csname bibitem#1\endcsname}%
\let\auto@bib@innerbib\@empty
\bibitem [{\citenamefont {Pickett}(2023)}]{pickett2023comment}%
  \BibitemOpen
  \bibfield  {author} {\bibinfo {author} {\bibfnamefont {W.~E.}\ \bibnamefont
  {Pickett}},\ }\href@noop {} {\bibinfo {title} {Comment on "nontrivial quantum
  geometry and the strength of electron-phonon coupling", arxiv:2305.02340, j.
  yu, c. j. ciccarino, r. bianco, i. errea, p. narang, b. a. bernevig}}
  (\bibinfo {year} {2023}),\ \Eprint {https://arxiv.org/abs/2306.02375}
  {arXiv:2306.02375 [cond-mat.supr-con]} \BibitemShut {NoStop}%
\bibitem [{\citenamefont {{Yu}}\ \emph {et~al.}(2023)\citenamefont {{Yu}},
  \citenamefont {{Ciccarino}}, \citenamefont {{Bianco}}, \citenamefont
  {{Errea}}, \citenamefont {{Narang}},\ and\ \citenamefont
  {{Bernevig}}}]{Yu05032023GeometryEPC}%
  \BibitemOpen
  \bibfield  {author} {\bibinfo {author} {\bibfnamefont {J.}~\bibnamefont
  {{Yu}}}, \bibinfo {author} {\bibfnamefont {C.~J.}\ \bibnamefont
  {{Ciccarino}}}, \bibinfo {author} {\bibfnamefont {R.}~\bibnamefont
  {{Bianco}}}, \bibinfo {author} {\bibfnamefont {I.}~\bibnamefont {{Errea}}},
  \bibinfo {author} {\bibfnamefont {P.}~\bibnamefont {{Narang}}},\ and\
  \bibinfo {author} {\bibfnamefont {B.~A.}\ \bibnamefont {{Bernevig}}},\
  }\bibfield  {title} {\bibinfo {title} {{Nontrivial Quantum Geometry and the
  Strength of Electron-Phonon Coupling}},\ }\href
  {https://doi.org/10.48550/arXiv.2305.02340} {\bibfield  {journal} {\bibinfo
  {journal} {arXiv e-prints}\ ,\ \bibinfo {eid} {arXiv:2305.02340}} (\bibinfo
  {year} {2023})},\ \Eprint {https://arxiv.org/abs/2305.02340}
  {arXiv:2305.02340 [cond-mat.supr-con]} \BibitemShut {NoStop}%
\bibitem [{\citenamefont {Provost}\ and\ \citenamefont
  {Vallee}(1980)}]{Provost1980FSMetric}%
  \BibitemOpen
  \bibfield  {author} {\bibinfo {author} {\bibfnamefont {J.}~\bibnamefont
  {Provost}}\ and\ \bibinfo {author} {\bibfnamefont {G.}~\bibnamefont
  {Vallee}},\ }\bibfield  {title} {\bibinfo {title} {Riemannian structure on
  manifolds of quantum states},\ }\href@noop {} {\bibfield  {journal} {\bibinfo
   {journal} {Communications in Mathematical Physics}\ }\textbf {\bibinfo
  {volume} {76}},\ \bibinfo {pages} {289} (\bibinfo {year} {1980})}\BibitemShut
  {NoStop}%
\bibitem [{\citenamefont {Resta}(2011)}]{Resta2011QuantumGeometry}%
  \BibitemOpen
  \bibfield  {author} {\bibinfo {author} {\bibfnamefont {R.}~\bibnamefont
  {Resta}},\ }\bibfield  {title} {\bibinfo {title} {The insulating state of
  matter: a geometrical theory},\ }\href
  {https://doi.org/10.1140/epjb/e2010-10874-4} {\bibfield  {journal} {\bibinfo
  {journal} {The European Physical Journal B}\ }\textbf {\bibinfo {volume}
  {79}},\ \bibinfo {pages} {121} (\bibinfo {year} {2011})}\BibitemShut
  {NoStop}%
\bibitem [{\citenamefont {Peotta}\ and\ \citenamefont
  {T{\"o}rm{\"a}}(2015)}]{Torma2015SWBoundChern}%
  \BibitemOpen
  \bibfield  {author} {\bibinfo {author} {\bibfnamefont {S.}~\bibnamefont
  {Peotta}}\ and\ \bibinfo {author} {\bibfnamefont {P.}~\bibnamefont
  {T{\"o}rm{\"a}}},\ }\bibfield  {title} {\bibinfo {title} {Superfluidity in
  topologically nontrivial flat bands},\ }\href
  {https://doi.org/10.1038/ncomms9944} {\bibfield  {journal} {\bibinfo
  {journal} {Nature Communications}\ }\textbf {\bibinfo {volume} {6}},\
  \bibinfo {pages} {8944} (\bibinfo {year} {2015})}\BibitemShut {NoStop}%
\bibitem [{\citenamefont {Julku}\ \emph {et~al.}(2016)\citenamefont {Julku},
  \citenamefont {Peotta}, \citenamefont {Vanhala}, \citenamefont {Kim},\ and\
  \citenamefont {T\"orm\"a}}]{Torma2016SuperfluidWeightLieb}%
  \BibitemOpen
  \bibfield  {author} {\bibinfo {author} {\bibfnamefont {A.}~\bibnamefont
  {Julku}}, \bibinfo {author} {\bibfnamefont {S.}~\bibnamefont {Peotta}},
  \bibinfo {author} {\bibfnamefont {T.~I.}\ \bibnamefont {Vanhala}}, \bibinfo
  {author} {\bibfnamefont {D.-H.}\ \bibnamefont {Kim}},\ and\ \bibinfo {author}
  {\bibfnamefont {P.}~\bibnamefont {T\"orm\"a}},\ }\bibfield  {title} {\bibinfo
  {title} {Geometric origin of superfluidity in the lieb-lattice flat band},\
  }\href {https://doi.org/10.1103/PhysRevLett.117.045303} {\bibfield  {journal}
  {\bibinfo  {journal} {Phys. Rev. Lett.}\ }\textbf {\bibinfo {volume} {117}},\
  \bibinfo {pages} {045303} (\bibinfo {year} {2016})}\BibitemShut {NoStop}%
\bibitem [{\citenamefont {T\"orm\"a}\ \emph {et~al.}(2018)\citenamefont
  {T\"orm\"a}, \citenamefont {Liang},\ and\ \citenamefont
  {Peotta}}]{Torma2018SelectiveQuantumMetric}%
  \BibitemOpen
  \bibfield  {author} {\bibinfo {author} {\bibfnamefont {P.}~\bibnamefont
  {T\"orm\"a}}, \bibinfo {author} {\bibfnamefont {L.}~\bibnamefont {Liang}},\
  and\ \bibinfo {author} {\bibfnamefont {S.}~\bibnamefont {Peotta}},\
  }\bibfield  {title} {\bibinfo {title} {Quantum metric and effective mass of a
  two-body bound state in a flat band},\ }\href
  {https://doi.org/10.1103/PhysRevB.98.220511} {\bibfield  {journal} {\bibinfo
  {journal} {Phys. Rev. B}\ }\textbf {\bibinfo {volume} {98}},\ \bibinfo
  {pages} {220511} (\bibinfo {year} {2018})}\BibitemShut {NoStop}%
\bibitem [{\citenamefont {Xie}\ \emph {et~al.}(2020)\citenamefont {Xie},
  \citenamefont {Song}, \citenamefont {Lian},\ and\ \citenamefont
  {Bernevig}}]{Xie2020TopologyBoundSCTBG}%
  \BibitemOpen
  \bibfield  {author} {\bibinfo {author} {\bibfnamefont {F.}~\bibnamefont
  {Xie}}, \bibinfo {author} {\bibfnamefont {Z.}~\bibnamefont {Song}}, \bibinfo
  {author} {\bibfnamefont {B.}~\bibnamefont {Lian}},\ and\ \bibinfo {author}
  {\bibfnamefont {B.~A.}\ \bibnamefont {Bernevig}},\ }\bibfield  {title}
  {\bibinfo {title} {Topology-bounded superfluid weight in twisted bilayer
  graphene},\ }\href {https://doi.org/10.1103/PhysRevLett.124.167002}
  {\bibfield  {journal} {\bibinfo  {journal} {Phys. Rev. Lett.}\ }\textbf
  {\bibinfo {volume} {124}},\ \bibinfo {pages} {167002} (\bibinfo {year}
  {2020})}\BibitemShut {NoStop}%
\bibitem [{\citenamefont {Herzog-Arbeitman}\ \emph {et~al.}(2021)\citenamefont
  {Herzog-Arbeitman}, \citenamefont {Peri}, \citenamefont {Schindler},
  \citenamefont {Huber},\ and\ \citenamefont
  {Bernevig}}]{Herzogarbeitman2021SWBound}%
  \BibitemOpen
  \bibfield  {author} {\bibinfo {author} {\bibfnamefont {J.}~\bibnamefont
  {Herzog-Arbeitman}}, \bibinfo {author} {\bibfnamefont {V.}~\bibnamefont
  {Peri}}, \bibinfo {author} {\bibfnamefont {F.}~\bibnamefont {Schindler}},
  \bibinfo {author} {\bibfnamefont {S.~D.}\ \bibnamefont {Huber}},\ and\
  \bibinfo {author} {\bibfnamefont {B.~A.}\ \bibnamefont {Bernevig}},\
  }\href@noop {} {\bibinfo {title} {Superfluid weight bounds from symmetry and
  quantum geometry in flat bands}} (\bibinfo {year} {2021}),\ \Eprint
  {https://arxiv.org/abs/2110.14663} {arXiv:2110.14663 [cond-mat.mes-hall]}
  \BibitemShut {NoStop}%
\bibitem [{\citenamefont {Verma}\ \emph {et~al.}(2021)\citenamefont {Verma},
  \citenamefont {Hazra},\ and\ \citenamefont {Randeria}}]{Verma2021FlatBandSC}%
  \BibitemOpen
  \bibfield  {author} {\bibinfo {author} {\bibfnamefont {N.}~\bibnamefont
  {Verma}}, \bibinfo {author} {\bibfnamefont {T.}~\bibnamefont {Hazra}},\ and\
  \bibinfo {author} {\bibfnamefont {M.}~\bibnamefont {Randeria}},\ }\bibfield
  {title} {\bibinfo {title} {Optical spectral weight, phase stiffness, and
  <i>t</i><sub><i>c</i></sub> bounds for trivial and topological flat band
  superconductors},\ }\href {https://doi.org/10.1073/pnas.2106744118}
  {\bibfield  {journal} {\bibinfo  {journal} {Proceedings of the National
  Academy of Sciences}\ }\textbf {\bibinfo {volume} {118}},\ \bibinfo {pages}
  {e2106744118} (\bibinfo {year} {2021})},\ \Eprint
  {https://arxiv.org/abs/https://www.pnas.org/doi/pdf/10.1073/pnas.2106744118}
  {https://www.pnas.org/doi/pdf/10.1073/pnas.2106744118} \BibitemShut {NoStop}%
\bibitem [{\citenamefont {Hu}\ \emph {et~al.}(2019)\citenamefont {Hu},
  \citenamefont {Hyart}, \citenamefont {Pikulin},\ and\ \citenamefont
  {Rossi}}]{Rossi2019SFWTBG}%
  \BibitemOpen
  \bibfield  {author} {\bibinfo {author} {\bibfnamefont {X.}~\bibnamefont
  {Hu}}, \bibinfo {author} {\bibfnamefont {T.}~\bibnamefont {Hyart}}, \bibinfo
  {author} {\bibfnamefont {D.~I.}\ \bibnamefont {Pikulin}},\ and\ \bibinfo
  {author} {\bibfnamefont {E.}~\bibnamefont {Rossi}},\ }\bibfield  {title}
  {\bibinfo {title} {Geometric and conventional contribution to the superfluid
  weight in twisted bilayer graphene},\ }\href
  {https://doi.org/10.1103/PhysRevLett.123.237002} {\bibfield  {journal}
  {\bibinfo  {journal} {Phys. Rev. Lett.}\ }\textbf {\bibinfo {volume} {123}},\
  \bibinfo {pages} {237002} (\bibinfo {year} {2019})}\BibitemShut {NoStop}%
\bibitem [{\citenamefont {Julku}\ \emph {et~al.}(2020)\citenamefont {Julku},
  \citenamefont {Peltonen}, \citenamefont {Liang}, \citenamefont {Heikkil\"a},\
  and\ \citenamefont {T\"orm\"a}}]{Torma2020SFWTBG}%
  \BibitemOpen
  \bibfield  {author} {\bibinfo {author} {\bibfnamefont {A.}~\bibnamefont
  {Julku}}, \bibinfo {author} {\bibfnamefont {T.~J.}\ \bibnamefont {Peltonen}},
  \bibinfo {author} {\bibfnamefont {L.}~\bibnamefont {Liang}}, \bibinfo
  {author} {\bibfnamefont {T.~T.}\ \bibnamefont {Heikkil\"a}},\ and\ \bibinfo
  {author} {\bibfnamefont {P.}~\bibnamefont {T\"orm\"a}},\ }\bibfield  {title}
  {\bibinfo {title} {Superfluid weight and berezinskii-kosterlitz-thouless
  transition temperature of twisted bilayer graphene},\ }\href
  {https://doi.org/10.1103/PhysRevB.101.060505} {\bibfield  {journal} {\bibinfo
   {journal} {Phys. Rev. B}\ }\textbf {\bibinfo {volume} {101}},\ \bibinfo
  {pages} {060505} (\bibinfo {year} {2020})}\BibitemShut {NoStop}%
\bibitem [{\citenamefont {Park}\ \emph {et~al.}(2020)\citenamefont {Park},
  \citenamefont {Kim},\ and\ \citenamefont {Lee}}]{Park2020SCHofBut}%
  \BibitemOpen
  \bibfield  {author} {\bibinfo {author} {\bibfnamefont {M.~J.}\ \bibnamefont
  {Park}}, \bibinfo {author} {\bibfnamefont {Y.~B.}\ \bibnamefont {Kim}},\ and\
  \bibinfo {author} {\bibfnamefont {S.}~\bibnamefont {Lee}},\ }\bibfield
  {title} {\bibinfo {title} {Geometric superconductivity in 3d hofstadter
  butterfly},\ }\href {https://arxiv.org/abs/2007.16205} {\bibfield  {journal}
  {\bibinfo  {journal} {arXiv:2007.16205}\ } (\bibinfo {year}
  {2020})}\BibitemShut {NoStop}%
\bibitem [{\citenamefont {Herzog-Arbeitman}\ \emph
  {et~al.}(2022{\natexlab{a}})\citenamefont {Herzog-Arbeitman}, \citenamefont
  {Peri}, \citenamefont {Schindler}, \citenamefont {Huber},\ and\ \citenamefont
  {Bernevig}}]{Herzog2022FlatBandQuantumGeometry}%
  \BibitemOpen
  \bibfield  {author} {\bibinfo {author} {\bibfnamefont {J.}~\bibnamefont
  {Herzog-Arbeitman}}, \bibinfo {author} {\bibfnamefont {V.}~\bibnamefont
  {Peri}}, \bibinfo {author} {\bibfnamefont {F.}~\bibnamefont {Schindler}},
  \bibinfo {author} {\bibfnamefont {S.~D.}\ \bibnamefont {Huber}},\ and\
  \bibinfo {author} {\bibfnamefont {B.~A.}\ \bibnamefont {Bernevig}},\
  }\bibfield  {title} {\bibinfo {title} {Superfluid weight bounds from symmetry
  and quantum geometry in flat bands},\ }\href
  {https://doi.org/10.1103/PhysRevLett.128.087002} {\bibfield  {journal}
  {\bibinfo  {journal} {Phys. Rev. Lett.}\ }\textbf {\bibinfo {volume} {128}},\
  \bibinfo {pages} {087002} (\bibinfo {year} {2022}{\natexlab{a}})}\BibitemShut
  {NoStop}%
\bibitem [{\citenamefont {Huhtinen}\ \emph {et~al.}(2022)\citenamefont
  {Huhtinen}, \citenamefont {Herzog-Arbeitman}, \citenamefont {Chew},
  \citenamefont {Bernevig},\ and\ \citenamefont
  {T\"orm\"a}}]{Huhtinen2022FlatBandSCQuantumMetric}%
  \BibitemOpen
  \bibfield  {author} {\bibinfo {author} {\bibfnamefont {K.-E.}\ \bibnamefont
  {Huhtinen}}, \bibinfo {author} {\bibfnamefont {J.}~\bibnamefont
  {Herzog-Arbeitman}}, \bibinfo {author} {\bibfnamefont {A.}~\bibnamefont
  {Chew}}, \bibinfo {author} {\bibfnamefont {B.~A.}\ \bibnamefont {Bernevig}},\
  and\ \bibinfo {author} {\bibfnamefont {P.}~\bibnamefont {T\"orm\"a}},\
  }\bibfield  {title} {\bibinfo {title} {Revisiting flat band
  superconductivity: Dependence on minimal quantum metric and band touchings},\
  }\href {https://doi.org/10.1103/PhysRevB.106.014518} {\bibfield  {journal}
  {\bibinfo  {journal} {Phys. Rev. B}\ }\textbf {\bibinfo {volume} {106}},\
  \bibinfo {pages} {014518} (\bibinfo {year} {2022})}\BibitemShut {NoStop}%
\bibitem [{\citenamefont {Yu}\ \emph {et~al.}(2022)\citenamefont {Yu},
  \citenamefont {Xie}, \citenamefont {Wu},\ and\ \citenamefont
  {Sarma}}]{Yu2022EOCPTBG}%
  \BibitemOpen
  \bibfield  {author} {\bibinfo {author} {\bibfnamefont {J.}~\bibnamefont
  {Yu}}, \bibinfo {author} {\bibfnamefont {M.}~\bibnamefont {Xie}}, \bibinfo
  {author} {\bibfnamefont {F.}~\bibnamefont {Wu}},\ and\ \bibinfo {author}
  {\bibfnamefont {S.~D.}\ \bibnamefont {Sarma}},\ }\href@noop {} {\bibinfo
  {title} {Euler obstructed cooper pairing in twisted bilayer graphene: Nematic
  nodal superconductivity and bounded superfluid weight}} (\bibinfo {year}
  {2022}),\ \Eprint {https://arxiv.org/abs/2202.02353} {arXiv:2202.02353
  [cond-mat.supr-con]} \BibitemShut {NoStop}%
\bibitem [{\citenamefont {Herzog-Arbeitman}\ \emph
  {et~al.}(2022{\natexlab{b}})\citenamefont {Herzog-Arbeitman}, \citenamefont
  {Chew}, \citenamefont {Huhtinen}, \citenamefont {T{\"o}rm{\"a}},\ and\
  \citenamefont {Bernevig}}]{Herzog2022ManyBodySCFlatBand}%
  \BibitemOpen
  \bibfield  {author} {\bibinfo {author} {\bibfnamefont {J.}~\bibnamefont
  {Herzog-Arbeitman}}, \bibinfo {author} {\bibfnamefont {A.}~\bibnamefont
  {Chew}}, \bibinfo {author} {\bibfnamefont {K.-E.}\ \bibnamefont {Huhtinen}},
  \bibinfo {author} {\bibfnamefont {P.}~\bibnamefont {T{\"o}rm{\"a}}},\ and\
  \bibinfo {author} {\bibfnamefont {B.~A.}\ \bibnamefont {Bernevig}},\
  }\bibfield  {title} {\bibinfo {title} {Many-body superconductivity in
  topological flat bands},\ }\href@noop {} {\bibfield  {journal} {\bibinfo
  {journal} {arXiv preprint arXiv:2209.00007}\ } (\bibinfo {year}
  {2022}{\natexlab{b}})}\BibitemShut {NoStop}%
\bibitem [{\citenamefont {{Chen}}\ and\ \citenamefont
  {{Huang}}(2022)}]{Huang2022arXivQuantumGeometryPDW}%
  \BibitemOpen
  \bibfield  {author} {\bibinfo {author} {\bibfnamefont {W.}~\bibnamefont
  {{Chen}}}\ and\ \bibinfo {author} {\bibfnamefont {W.}~\bibnamefont
  {{Huang}}},\ }\bibfield  {title} {\bibinfo {title} {{Pair density wave
  facilitated by Bloch quantum geometry in nearly flat band multiorbital
  superconductors}},\ }\href {https://doi.org/10.48550/arXiv.2208.02285}
  {\bibfield  {journal} {\bibinfo  {journal} {arXiv e-prints}\ ,\ \bibinfo
  {eid} {arXiv:2208.02285}} (\bibinfo {year} {2022})},\ \Eprint
  {https://arxiv.org/abs/2208.02285} {arXiv:2208.02285 [cond-mat.supr-con]}
  \BibitemShut {NoStop}%
\bibitem [{\citenamefont {Chen}\ and\ \citenamefont
  {Law}(2023)}]{Law2023QuantumMetricLandau}%
  \BibitemOpen
  \bibfield  {author} {\bibinfo {author} {\bibfnamefont {S.~A.}\ \bibnamefont
  {Chen}}\ and\ \bibinfo {author} {\bibfnamefont {K.}~\bibnamefont {Law}},\
  }\bibfield  {title} {\bibinfo {title} {Towards a ginzburg-landau theory of
  the quantum geometric effect in superconductors},\ }\href@noop {} {\bibfield
  {journal} {\bibinfo  {journal} {arXiv preprint arXiv:2303.15504}\ } (\bibinfo
  {year} {2023})}\BibitemShut {NoStop}%
\bibitem [{\citenamefont {T{\"o}rm{\"a}}\ \emph {et~al.}(2022)\citenamefont
  {T{\"o}rm{\"a}}, \citenamefont {Peotta},\ and\ \citenamefont
  {Bernevig}}]{Torma2022ReviewQuantumGeometry}%
  \BibitemOpen
  \bibfield  {author} {\bibinfo {author} {\bibfnamefont {P.}~\bibnamefont
  {T{\"o}rm{\"a}}}, \bibinfo {author} {\bibfnamefont {S.}~\bibnamefont
  {Peotta}},\ and\ \bibinfo {author} {\bibfnamefont {B.~A.}\ \bibnamefont
  {Bernevig}},\ }\bibfield  {title} {\bibinfo {title} {Superconductivity,
  superfluidity and quantum geometry in twisted multilayer systems},\ }\href
  {https://doi.org/10.1038/s42254-022-00466-y} {\bibfield  {journal} {\bibinfo
  {journal} {Nature Reviews Physics}\ }\textbf {\bibinfo {volume} {4}},\
  \bibinfo {pages} {528} (\bibinfo {year} {2022})}\BibitemShut {NoStop}%
\bibitem [{\citenamefont {{Hofmann}}\ \emph {et~al.}(2022)\citenamefont
  {{Hofmann}}, \citenamefont {{Berg}},\ and\ \citenamefont
  {{Chowdhury}}}]{Chowdhury2022SFFlatBandQuantumGeometry}%
  \BibitemOpen
  \bibfield  {author} {\bibinfo {author} {\bibfnamefont {J.~S.}\ \bibnamefont
  {{Hofmann}}}, \bibinfo {author} {\bibfnamefont {E.}~\bibnamefont {{Berg}}},\
  and\ \bibinfo {author} {\bibfnamefont {D.}~\bibnamefont {{Chowdhury}}},\
  }\bibfield  {title} {\bibinfo {title} {{Superconductivity, charge density
  wave, and supersolidity in flat bands with tunable quantum metric}},\ }\href
  {https://doi.org/10.48550/arXiv.2204.02994} {\bibfield  {journal} {\bibinfo
  {journal} {arXiv e-prints}\ ,\ \bibinfo {eid} {arXiv:2204.02994}} (\bibinfo
  {year} {2022})},\ \Eprint {https://arxiv.org/abs/2204.02994}
  {arXiv:2204.02994 [cond-mat.str-el]} \BibitemShut {NoStop}%
\bibitem [{\citenamefont {Regnault}\ and\ \citenamefont
  {Bernevig}(2011)}]{BAB2011FCI}%
  \BibitemOpen
  \bibfield  {author} {\bibinfo {author} {\bibfnamefont {N.}~\bibnamefont
  {Regnault}}\ and\ \bibinfo {author} {\bibfnamefont {B.~A.}\ \bibnamefont
  {Bernevig}},\ }\bibfield  {title} {\bibinfo {title} {Fractional chern
  insulator},\ }\href {https://doi.org/10.1103/PhysRevX.1.021014} {\bibfield
  {journal} {\bibinfo  {journal} {Phys. Rev. X}\ }\textbf {\bibinfo {volume}
  {1}},\ \bibinfo {pages} {021014} (\bibinfo {year} {2011})}\BibitemShut
  {NoStop}%
\bibitem [{\citenamefont {Parameswaran}\ \emph {et~al.}(2013)\citenamefont
  {Parameswaran}, \citenamefont {Roy},\ and\ \citenamefont
  {Sondhi}}]{Sondhi2013FCI}%
  \BibitemOpen
  \bibfield  {author} {\bibinfo {author} {\bibfnamefont {S.~A.}\ \bibnamefont
  {Parameswaran}}, \bibinfo {author} {\bibfnamefont {R.}~\bibnamefont {Roy}},\
  and\ \bibinfo {author} {\bibfnamefont {S.~L.}\ \bibnamefont {Sondhi}},\
  }\bibfield  {title} {\bibinfo {title} {Fractional quantum hall physics in
  topological flat bands},\ }\href
  {https://doi.org/https://doi.org/10.1016/j.crhy.2013.04.003} {\bibfield
  {journal} {\bibinfo  {journal} {Comptes Rendus Physique}\ }\textbf {\bibinfo
  {volume} {14}},\ \bibinfo {pages} {816} (\bibinfo {year} {2013})},\ \bibinfo
  {note} {topological insulators / Isolants topologiques}\BibitemShut {NoStop}%
\bibitem [{\citenamefont {Dobard\ifmmode \check{z}\else
  \v{z}\fi{}i\ifmmode~\acute{c}\else \'{c}\fi{}}\ \emph
  {et~al.}(2013)\citenamefont {Dobard\ifmmode \check{z}\else
  \v{z}\fi{}i\ifmmode~\acute{c}\else \'{c}\fi{}}, \citenamefont
  {Milovanovi\ifmmode~\acute{c}\else \'{c}\fi{}},\ and\ \citenamefont
  {Regnault}}]{Regnault2013FCI}%
  \BibitemOpen
  \bibfield  {author} {\bibinfo {author} {\bibfnamefont {E.}~\bibnamefont
  {Dobard\ifmmode \check{z}\else \v{z}\fi{}i\ifmmode~\acute{c}\else
  \'{c}\fi{}}}, \bibinfo {author} {\bibfnamefont {M.~V.}\ \bibnamefont
  {Milovanovi\ifmmode~\acute{c}\else \'{c}\fi{}}},\ and\ \bibinfo {author}
  {\bibfnamefont {N.}~\bibnamefont {Regnault}},\ }\bibfield  {title} {\bibinfo
  {title} {Geometrical description of fractional chern insulators based on
  static structure factor calculations},\ }\href
  {https://doi.org/10.1103/PhysRevB.88.115117} {\bibfield  {journal} {\bibinfo
  {journal} {Phys. Rev. B}\ }\textbf {\bibinfo {volume} {88}},\ \bibinfo
  {pages} {115117} (\bibinfo {year} {2013})}\BibitemShut {NoStop}%
\bibitem [{\citenamefont {Roy}(2014)}]{Roy2014FCI}%
  \BibitemOpen
  \bibfield  {author} {\bibinfo {author} {\bibfnamefont {R.}~\bibnamefont
  {Roy}},\ }\bibfield  {title} {\bibinfo {title} {Band geometry of fractional
  topological insulators},\ }\href {https://doi.org/10.1103/PhysRevB.90.165139}
  {\bibfield  {journal} {\bibinfo  {journal} {Phys. Rev. B}\ }\textbf {\bibinfo
  {volume} {90}},\ \bibinfo {pages} {165139} (\bibinfo {year}
  {2014})}\BibitemShut {NoStop}%
\bibitem [{\citenamefont {Ledwith}\ \emph {et~al.}(2020)\citenamefont
  {Ledwith}, \citenamefont {Tarnopolsky}, \citenamefont {Khalaf},\ and\
  \citenamefont {Vishwanath}}]{Vishwanath2020FCITBG}%
  \BibitemOpen
  \bibfield  {author} {\bibinfo {author} {\bibfnamefont {P.~J.}\ \bibnamefont
  {Ledwith}}, \bibinfo {author} {\bibfnamefont {G.}~\bibnamefont
  {Tarnopolsky}}, \bibinfo {author} {\bibfnamefont {E.}~\bibnamefont
  {Khalaf}},\ and\ \bibinfo {author} {\bibfnamefont {A.}~\bibnamefont
  {Vishwanath}},\ }\bibfield  {title} {\bibinfo {title} {Fractional chern
  insulator states in twisted bilayer graphene: An analytical approach},\
  }\href {https://doi.org/10.1103/PhysRevResearch.2.023237} {\bibfield
  {journal} {\bibinfo  {journal} {Phys. Rev. Res.}\ }\textbf {\bibinfo {volume}
  {2}},\ \bibinfo {pages} {023237} (\bibinfo {year} {2020})}\BibitemShut
  {NoStop}%
\bibitem [{\citenamefont {Wang}\ and\ \citenamefont
  {Liu}(2022)}]{Wang2022FCITwistedGraphene}%
  \BibitemOpen
  \bibfield  {author} {\bibinfo {author} {\bibfnamefont {J.}~\bibnamefont
  {Wang}}\ and\ \bibinfo {author} {\bibfnamefont {Z.}~\bibnamefont {Liu}},\
  }\bibfield  {title} {\bibinfo {title} {Hierarchy of ideal flatbands in chiral
  twisted multilayer graphene models},\ }\href
  {https://doi.org/10.1103/PhysRevLett.128.176403} {\bibfield  {journal}
  {\bibinfo  {journal} {Phys. Rev. Lett.}\ }\textbf {\bibinfo {volume} {128}},\
  \bibinfo {pages} {176403} (\bibinfo {year} {2022})}\BibitemShut {NoStop}%
\bibitem [{\citenamefont {Pi\'echon}\ \emph {et~al.}(2016)\citenamefont
  {Pi\'echon}, \citenamefont {Raoux}, \citenamefont {Fuchs},\ and\
  \citenamefont {Montambaux}}]{Montambaux2016GeometryObitalSusceptibility}%
  \BibitemOpen
  \bibfield  {author} {\bibinfo {author} {\bibfnamefont {F.}~\bibnamefont
  {Pi\'echon}}, \bibinfo {author} {\bibfnamefont {A.}~\bibnamefont {Raoux}},
  \bibinfo {author} {\bibfnamefont {J.-N.}\ \bibnamefont {Fuchs}},\ and\
  \bibinfo {author} {\bibfnamefont {G.}~\bibnamefont {Montambaux}},\ }\bibfield
   {title} {\bibinfo {title} {Geometric orbital susceptibility: Quantum metric
  without berry curvature},\ }\href
  {https://doi.org/10.1103/PhysRevB.94.134423} {\bibfield  {journal} {\bibinfo
  {journal} {Phys. Rev. B}\ }\textbf {\bibinfo {volume} {94}},\ \bibinfo
  {pages} {134423} (\bibinfo {year} {2016})}\BibitemShut {NoStop}%
\bibitem [{\citenamefont {Rhim}\ \emph {et~al.}(2020)\citenamefont {Rhim},
  \citenamefont {Kim},\ and\ \citenamefont
  {Yang}}]{Yang2020QuantumDistanceFlatBands}%
  \BibitemOpen
  \bibfield  {author} {\bibinfo {author} {\bibfnamefont {J.-W.}\ \bibnamefont
  {Rhim}}, \bibinfo {author} {\bibfnamefont {K.}~\bibnamefont {Kim}},\ and\
  \bibinfo {author} {\bibfnamefont {B.-J.}\ \bibnamefont {Yang}},\ }\bibfield
  {title} {\bibinfo {title} {Quantum distance and anomalous landau levels of
  flat bands},\ }\href {https://doi.org/10.1038/s41586-020-2540-1} {\bibfield
  {journal} {\bibinfo  {journal} {Nature}\ }\textbf {\bibinfo {volume} {584}},\
  \bibinfo {pages} {59} (\bibinfo {year} {2020})}\BibitemShut {NoStop}%
\bibitem [{\citenamefont {Mera}\ and\ \citenamefont
  {Ozawa}(2021)}]{Mera2021FlatBandsKahler}%
  \BibitemOpen
  \bibfield  {author} {\bibinfo {author} {\bibfnamefont {B.}~\bibnamefont
  {Mera}}\ and\ \bibinfo {author} {\bibfnamefont {T.}~\bibnamefont {Ozawa}},\
  }\bibfield  {title} {\bibinfo {title} {Engineering geometrically flat chern
  bands with fubini-study k\"ahler structure},\ }\href
  {https://doi.org/10.1103/PhysRevB.104.115160} {\bibfield  {journal} {\bibinfo
   {journal} {Phys. Rev. B}\ }\textbf {\bibinfo {volume} {104}},\ \bibinfo
  {pages} {115160} (\bibinfo {year} {2021})}\BibitemShut {NoStop}%
\bibitem [{\citenamefont {Julku}\ \emph {et~al.}(2021)\citenamefont {Julku},
  \citenamefont {Bruun},\ and\ \citenamefont
  {T\"orm\"a}}]{Torma2021FlatBandBEC}%
  \BibitemOpen
  \bibfield  {author} {\bibinfo {author} {\bibfnamefont {A.}~\bibnamefont
  {Julku}}, \bibinfo {author} {\bibfnamefont {G.~M.}\ \bibnamefont {Bruun}},\
  and\ \bibinfo {author} {\bibfnamefont {P.}~\bibnamefont {T\"orm\"a}},\
  }\bibfield  {title} {\bibinfo {title} {Quantum geometry and flat band
  bose-einstein condensation},\ }\href
  {https://doi.org/10.1103/PhysRevLett.127.170404} {\bibfield  {journal}
  {\bibinfo  {journal} {Phys. Rev. Lett.}\ }\textbf {\bibinfo {volume} {127}},\
  \bibinfo {pages} {170404} (\bibinfo {year} {2021})}\BibitemShut {NoStop}%
\bibitem [{\citenamefont {Wang}\ \emph {et~al.}(2021)\citenamefont {Wang},
  \citenamefont {Cano}, \citenamefont {Millis}, \citenamefont {Liu},\ and\
  \citenamefont {Yang}}]{Wang2021GeometryFlatBand}%
  \BibitemOpen
  \bibfield  {author} {\bibinfo {author} {\bibfnamefont {J.}~\bibnamefont
  {Wang}}, \bibinfo {author} {\bibfnamefont {J.}~\bibnamefont {Cano}}, \bibinfo
  {author} {\bibfnamefont {A.~J.}\ \bibnamefont {Millis}}, \bibinfo {author}
  {\bibfnamefont {Z.}~\bibnamefont {Liu}},\ and\ \bibinfo {author}
  {\bibfnamefont {B.}~\bibnamefont {Yang}},\ }\bibfield  {title} {\bibinfo
  {title} {Exact landau level description of geometry and interaction in a
  flatband},\ }\href {https://doi.org/10.1103/PhysRevLett.127.246403}
  {\bibfield  {journal} {\bibinfo  {journal} {Phys. Rev. Lett.}\ }\textbf
  {\bibinfo {volume} {127}},\ \bibinfo {pages} {246403} (\bibinfo {year}
  {2021})}\BibitemShut {NoStop}%
\bibitem [{\citenamefont {Hu}\ \emph {et~al.}(2022)\citenamefont {Hu},
  \citenamefont {Hyart}, \citenamefont {Pikulin},\ and\ \citenamefont
  {Rossi}}]{Rossi2022QuantumMetricExciton}%
  \BibitemOpen
  \bibfield  {author} {\bibinfo {author} {\bibfnamefont {X.}~\bibnamefont
  {Hu}}, \bibinfo {author} {\bibfnamefont {T.}~\bibnamefont {Hyart}}, \bibinfo
  {author} {\bibfnamefont {D.~I.}\ \bibnamefont {Pikulin}},\ and\ \bibinfo
  {author} {\bibfnamefont {E.}~\bibnamefont {Rossi}},\ }\bibfield  {title}
  {\bibinfo {title} {Quantum-metric-enabled exciton condensate in double
  twisted bilayer graphene},\ }\href
  {https://doi.org/10.1103/PhysRevB.105.L140506} {\bibfield  {journal}
  {\bibinfo  {journal} {Phys. Rev. B}\ }\textbf {\bibinfo {volume} {105}},\
  \bibinfo {pages} {L140506} (\bibinfo {year} {2022})}\BibitemShut {NoStop}%
\bibitem [{\citenamefont {Mitscherling}\ and\ \citenamefont
  {Holder}(2022)}]{Holder2022FlatBandQuantumMetricResistivity}%
  \BibitemOpen
  \bibfield  {author} {\bibinfo {author} {\bibfnamefont {J.}~\bibnamefont
  {Mitscherling}}\ and\ \bibinfo {author} {\bibfnamefont {T.}~\bibnamefont
  {Holder}},\ }\bibfield  {title} {\bibinfo {title} {Bound on resistivity in
  flat-band materials due to the quantum metric},\ }\href
  {https://doi.org/10.1103/PhysRevB.105.085154} {\bibfield  {journal} {\bibinfo
   {journal} {Phys. Rev. B}\ }\textbf {\bibinfo {volume} {105}},\ \bibinfo
  {pages} {085154} (\bibinfo {year} {2022})}\BibitemShut {NoStop}%
\bibitem [{\citenamefont {Chaudhary}\ \emph {et~al.}(2022)\citenamefont
  {Chaudhary}, \citenamefont {Lewandowski},\ and\ \citenamefont
  {Refael}}]{Refael2022ShiftCurrentTBG}%
  \BibitemOpen
  \bibfield  {author} {\bibinfo {author} {\bibfnamefont {S.}~\bibnamefont
  {Chaudhary}}, \bibinfo {author} {\bibfnamefont {C.}~\bibnamefont
  {Lewandowski}},\ and\ \bibinfo {author} {\bibfnamefont {G.}~\bibnamefont
  {Refael}},\ }\bibfield  {title} {\bibinfo {title} {Shift-current response as
  a probe of quantum geometry and electron-electron interactions in twisted
  bilayer graphene},\ }\href {https://doi.org/10.1103/PhysRevResearch.4.013164}
  {\bibfield  {journal} {\bibinfo  {journal} {Phys. Rev. Res.}\ }\textbf
  {\bibinfo {volume} {4}},\ \bibinfo {pages} {013164} (\bibinfo {year}
  {2022})}\BibitemShut {NoStop}%
\bibitem [{\citenamefont {Oh}\ \emph {et~al.}(2022)\citenamefont {Oh},
  \citenamefont {Cho}, \citenamefont {Park},\ and\ \citenamefont
  {Rhim}}]{Oh04252022RhimQuantumDistance}%
  \BibitemOpen
  \bibfield  {author} {\bibinfo {author} {\bibfnamefont {C.-g.}\ \bibnamefont
  {Oh}}, \bibinfo {author} {\bibfnamefont {D.}~\bibnamefont {Cho}}, \bibinfo
  {author} {\bibfnamefont {S.~Y.}\ \bibnamefont {Park}},\ and\ \bibinfo
  {author} {\bibfnamefont {J.-W.}\ \bibnamefont {Rhim}},\ }\bibfield  {title}
  {\bibinfo {title} {Bulk-interface correspondence from quantum distance in
  flat band systems},\ }\href {https://doi.org/10.1038/s42005-022-01102-y}
  {\bibfield  {journal} {\bibinfo  {journal} {Communications Physics}\ }\textbf
  {\bibinfo {volume} {5}},\ \bibinfo {pages} {320} (\bibinfo {year}
  {2022})}\BibitemShut {NoStop}%
\bibitem [{\citenamefont {Fan}\ \emph {et~al.}(2002)\citenamefont {Fan},
  \citenamefont {Ru-Shan}, \citenamefont {Ning-Hua},\ and\ \citenamefont
  {Wei}}]{Fan03052002MgB2}%
  \BibitemOpen
  \bibfield  {author} {\bibinfo {author} {\bibfnamefont {Y.}~\bibnamefont
  {Fan}}, \bibinfo {author} {\bibfnamefont {H.}~\bibnamefont {Ru-Shan}},
  \bibinfo {author} {\bibfnamefont {T.}~\bibnamefont {Ning-Hua}},\ and\
  \bibinfo {author} {\bibfnamefont {G.}~\bibnamefont {Wei}},\ }\bibfield
  {title} {\bibinfo {title} {Electronic structural properties and
  superconductivity of diborides in the mgb2 structure},\ }\href@noop {}
  {\bibfield  {journal} {\bibinfo  {journal} {Chinese physics letters}\
  }\textbf {\bibinfo {volume} {19}},\ \bibinfo {pages} {1336} (\bibinfo {year}
  {2002})}\BibitemShut {NoStop}%
\bibitem [{\citenamefont {An}\ and\ \citenamefont
  {Pickett}(2001)}]{Pickett2011CovalentBondsDriven}%
  \BibitemOpen
  \bibfield  {author} {\bibinfo {author} {\bibfnamefont {J.~M.}\ \bibnamefont
  {An}}\ and\ \bibinfo {author} {\bibfnamefont {W.~E.}\ \bibnamefont
  {Pickett}},\ }\bibfield  {title} {\bibinfo {title} {Superconductivity of
  ${\mathrm{mgb}}_{2}$: Covalent bonds driven metallic},\ }\href
  {https://doi.org/10.1103/PhysRevLett.86.4366} {\bibfield  {journal} {\bibinfo
   {journal} {Phys. Rev. Lett.}\ }\textbf {\bibinfo {volume} {86}},\ \bibinfo
  {pages} {4366} (\bibinfo {year} {2001})}\BibitemShut {NoStop}%
\end{thebibliography}%

\appendix

\clearpage

\tableofcontents

\section{The particular importance of the $\sigma$-bonding states and the $E_{2g}$ phonon modes in {\mgb} is explicitly manifested in our results derived from our EPC model.}

{\pcomm} claims that our work (especially our EPC model) has ``no hint of the colossal impact of the cylindrical Fermi surfaces and the uniquely large EPC matrix elements involving the electron states on the Fermi surface and the scattering processes across these Fermi surfaces."
We respectfully disagree with this claim.
First, the cylindrical Fermi surfaces in {\mgb} are given by the $\sigma$-bonding electron states.
The dominance (or colossal impact) of the cylindrical Fermi surfaces or the $\sigma$-bonding electron states in the EPC constant $\lambda$ is explicitly manifested in our result derived from our EPC model: the EPC constant $\lambda$ of {\mgb} from our EPC model is dominated by the geometric contribution $\lambda_{\sigma,geo}$ from the $\sigma$-bonding electron states (or from the integral on the cylindrical Fermi surfaces), as shown in TABLE I of our work~\cite{Yu05032023GeometryEPC}.
Second, the uniquely large EPC matrix elements involving the electron states on the cylindrical Fermi surfaces and the scattering processes across these Fermi surfaces are caused by the $E_{2g}$ bond-stretching phonon modes.
The dominance of the $E_{2g}$ bond-stretching phonon modes is explicitly shown in the final expression of the dominant $\sigma$-bonding geometric contribution $\lambda_{\sigma,geo}$, as the orbital-selective Fubini-Study metric in $\lambda_{\sigma,geo}$ is exactly selected by the $E_{2g}$ bond-stretching phonon modes (Eq.\,(7-8) of our work~\cite{Yu05032023GeometryEPC}), \ie, only the $E_{2g}$ phonons have contributions to $\lambda_{\sigma,geo}$ under our leading-order approximation.
If tracking the derivation of the $\lambda_{\sigma,geo}$ in App.\,H4d, it is clear that the EPC matrix elements or the scattering processes between $\sigma$-bonding states from the $E_{2g}$ bond-stretching phonon modes are included in Eq.\,(H205) of our work~\cite{Yu05032023GeometryEPC}.
As a result, the colossal impact of the cylindrical Fermi surfaces, as well as the uniquely large EPC matrix elements or scattering processes among them, are naturally and explictly manifested in our final results from our EPC model.

In addition, {\pcomm} claims that our work (especially our EPC model that is built from the hopping change) did not emphasize or account for ``the essence of the 3\% of very strongly coupled phonons", where the $3\%$ of the very strongly coupled phonons are basically $E_{2g}$ phonons (or more precisely $E_2$ phonons along $\Gamma$-A) and phonons that are similar to them.
This claim is also falsified by the fact that the orbital-selective Fubini-Study metric in dominant $\lambda_{\sigma,geo}$ derived from our EPC model is exactly selected by the $E_{2g}$ bond-stretching phonon modes (Eq.\,(7-8) of our work~\cite{Yu05032023GeometryEPC})---which naturally show the particular importance of the $E_{2g}$ bond-stretching phonon modes. 
{\pcomm} also claims that ``the average of the electron-ion quantity $\Gamma_{nn'}(\bsl{k},\bsl{k}')$ seems to incur no unusual physics, as
phonon-related effect have been neglected", where $\Gamma_{nn'}(\bsl{k},\bsl{k}')$ is defined in Eq.\,(1) of our work~\cite{Yu05032023GeometryEPC}.
This claim is falsified by the fact that the $E_{2g}$-phonon-selected orbital-selective Fubini-Study metric is derived from the average of $\Gamma_{nn'}(\bsl{k},\bsl{k}')$ (\ie, $\langle \Gamma \rangle$), demonstrating the phonon-related effect in $\langle \Gamma \rangle$.
In other words, the results derived from our EPC model naturally and explicitly show the dominance of the $E_{2g}$ bond-stretching phonon modes or the strong $B-B$ bond in $\langle \Gamma \rangle$ of $\lambda$ in addition to $\langle \omega^2 \rangle$, where $\langle \omega^2 \rangle$ is the mean-squared phonon frequency.

\section{The bond-bending effect has been included in our derivation.}

{\pcomm} claims that our work used a hopping change model for {\mgb} that neglects the bond-bending effect (for the dominant $\sigma$ bond) in the EPC.
We respectfully disagree with this claim, since the bond-bending effect has been explicitly included in our derivation, as explicitly shown in App.\,H4 of our work~\cite{Yu05032023GeometryEPC} (more specifically the second term of the second last line of Eq.\,(H160)).
Tracking back the derivation, we have explicitly included the the angular dependence of the $p_x$ and $p_y$ orbitals as shown in Eq.\,(H152) of our work~\cite{Yu05032023GeometryEPC}, which eventually leads to the bond-bending effect.
Nevertheless, the bond-bending term in Eq.\,(H160) eventually cancels with the last term in Eq.\,(H160) under the nearest-neighboring-hopping and first-order-$k_x/k_y$ approximations of the electron Hamiltonian, which allows us to get the simple Gaussian form of the EPC in Eq.(H173), since the two terms that cancel are exactly those that do not directly come from the Gaussian approximation.
In the symmetry-representation method, the bond-bending effect has also been included in App.\,H5a of our work (specifically the $\hat{\gamma}_7$ term in Eq.\,(H247)), which comes from $\widetilde{f}_{\bsl{\tau}_1\bsl{\tau}_2,\perp}$ in Eq.\,(H61) in App.\,H2 of our work.
The final result of the symmetry-representation method is consistent with that from the Gaussian approximation, under the nearest-neighboring-hopping and first-order-$k_x/k_y$ approximations of the $\sigma$-bonding electron Hamiltonian.
Certainly, if we go beyond the first-order-$k_x/k_y$-approximation of the $\sigma$-bonding electron Hamiltonian, the bond-bending term will explicitly appear in the final expression of the EPC Hamiltonian, and give rise to contributions to the energetic/geometric parts of the EPC, which will make the expression of the geometric contribution more complicated but meanwhile may improve the match between EPC model and the {\abi} calculation.
This would be an interesting future work.

\section{Density response (in the context of {\abi} calculation) has been included in our model.}

\begin{figure*}[t]
    \centering
    \includegraphics[width=1.6\columnwidth]{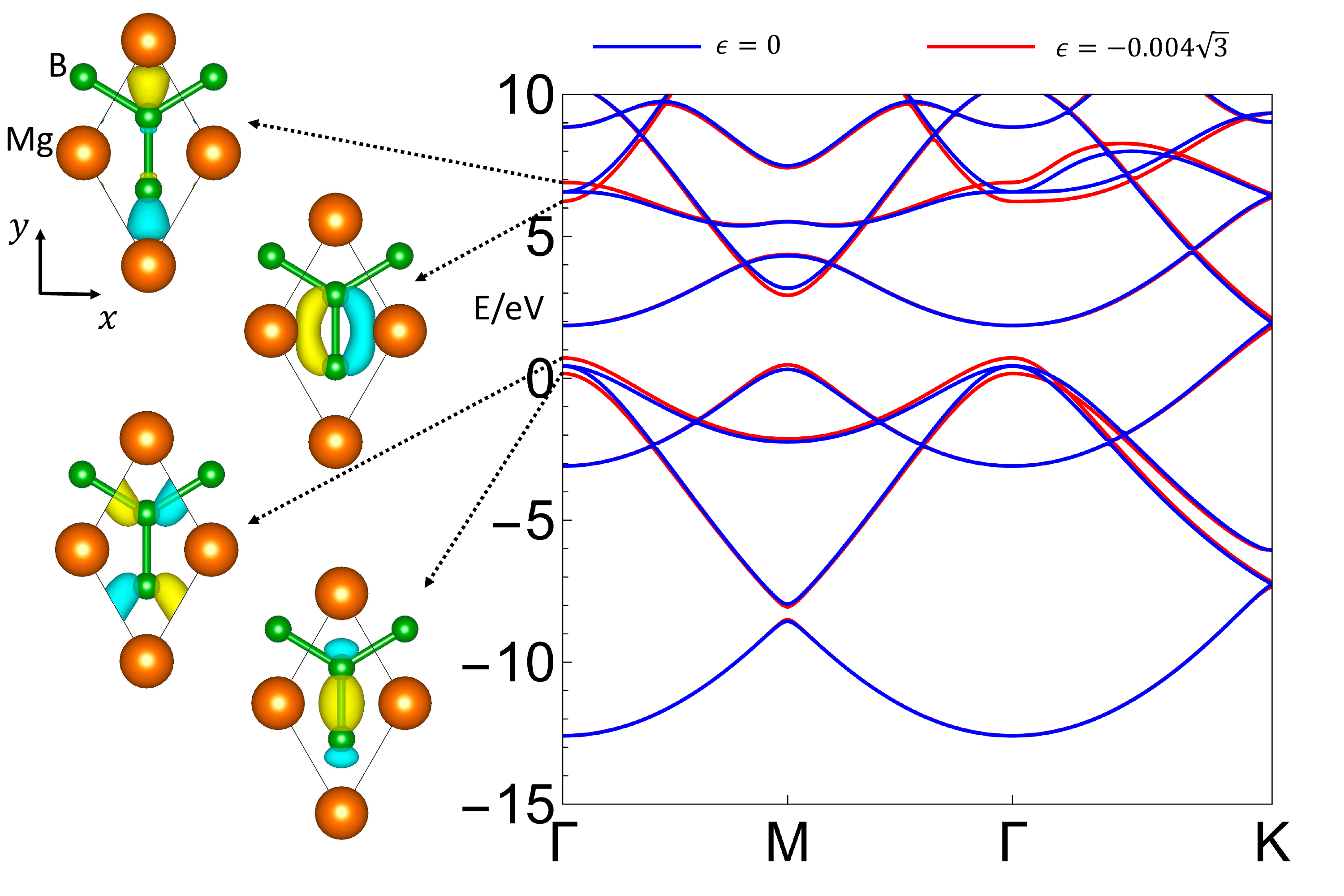}
    \caption{The band structures (right) for {\mgb} with no frozen phonons (blue) and with a frozen $E_{2g}$ bond-stretching mode along $y$.
    The chemical potential is at zero.
    The change of the bond length for the bond-stretching mode is $ 2 \epsilon a = -0.008 \sqrt{3} a$ with $a$ being the in-plane lattice constant and $\epsilon$ measures the bond length change.
    $\Gamma-M$ is along $y$ and $\Gamma-K$ is along $x$, and thus both paths have mirror symmetries.
    The 4 plots on the left are the 4 $p_x/p_y$ wave-functions at $\Gamma$ for {\mgb} with frozen $E_{2g}$ bond stretching with $\epsilon = -0.004 \sqrt{3}$.
    The blue and yellow parts of the wavefunction have opposite signs. 
    }
    \label{fig:MgB2_bands_frozen_E2g_phonons}
\end{figure*}

{\pcomm} claims that our work has not considered the electron density response to the ion motions, which is an essential part of the {\abi} calculation of the EPC.
Here the density response mentioned in {\pcomm} refers to the electron density change caused by ion motions in the context of the {\abi} calculation, instead of the Wannier-onsite EPC in the sense of, \eg, Holstein model.
We respectfully disagree with the claim in {\pcomm}, because our EPC model has included the density response in the {\abi} calculation.
The {\abi} calculation of the EPC is done by calculating the change of the Kohn-Sham potential due to ion motions (which includes the density response).
In our work, we use the hopping change to model the change of the Kohn-Sham potential projected to the atomic Wannier functions, where the Wannier projection does not lose any information relevant to the low-energy physics, and the parameters in our model are determined from the {\abi} calculation.
Crucially, our EPC model built from the hopping change (or more generally from two-center approximation) does not miss any considerable terms in the EPC given by the change of the Kohn-Sham potential in {\abi} calculation, as shown by the comparison of (gauge-invariant combinations of) the EPC matrix elements in Fig.\,9 of our work~\cite{Yu05032023GeometryEPC} and by the discussion below on the small Wannier-onsite EPC.
Therefore, our EPC model from the hopping change is a good model for the ion-motion-induced change of the Kohn-Sham potential projected to the atomic Wannier functions, and thus has included the density response.

As mentioned in the above paragraph, one EPC term neglected in our model is the Wannier-onsite EPC, which comes from the projection of the change of Kohn-Sham potential to the two atomic (electron) Wannier functions at the same site.
Nevertheless, as implied by the $\Gamma-M$ bands of Fig.\,2 of \refcite{Fan03052002MgB2}, the Wannier-onsite EPC is indeed negligible, further justifying our model.
In the following, we will present a more thorough discussion on this point.
From the {\abi} calculation, it is known the EPC in {\mgb} is dominated by the $E_{2g}$ bond-stretching modes~\cite{Pickett2011CovalentBondsDriven}.
In principle, the EPC for the $E_{2g}$ bond-stretching modes in the basis of the atomic Wannier functions (or atomic orbitals) might have two origins: (i) the change of the electron Wannier-onsite terms due to the bond-stretching modes (onsite EPC), and (ii) the electron hopping change due to the bond-stretching modes (hopping-change EPC).
To show the Wannier-onsite EPC is negligible, let us consider a frozen $E_{2g}$ bond-stretching phonon at $\Gamma$, which has been used to show the large coupling between $E_{2g}$ phonons and electrons in {\mgb} by looking at the gap opened between two $p_x/p_y$-parity-even states (degenerate without the frozen $E_{2g}$ bond-stretching phonons) near the Fermi level~\cite{Pickett2011CovalentBondsDriven}.
In fact, the $E_{2g}$ frozen phonon will also open a gap between two previously-degenerate $p_x/p_y$-parity states that are far away from the Fermi energy (about $6$eV above the Fermi level)~\cite{Fan03052002MgB2}.
Without loss of generality, let us consider the bond-stretching mode along $y$; our {\abi} calculations show that the $p_y$ bonding and anti-bonding states have opposite energy shifts, so do the $p_x$ bonding and anti-bonding states, as explicitly shown in \figref{fig:MgB2_bands_frozen_E2g_phonons}.
(The $\Gamma-M$ bands of Fig.\,2 of \refcite{Fan03052002MgB2} have the same band crossing features as those in \figref{fig:MgB2_bands_frozen_E2g_phonons}, which are signatures of the opposite energy shifts.)
Such opposite energy shifts for the bonding and anti-bonding states of the same type of orbitals are consistent with the dominance of the hopping change, instead of the Wannier-onsite EPC which would give the same energy shifts for the bonding and anti-bonding states of the same type of orbitals.
Explicit comparison based on the {\abi} data shows the ratio between the Wannier-onsite EPC and the hopping-change EPC is less 0.1, which means the Wannier-onsite EPC can be neglected.

\section{Our results do not rely on Gaussian approximation.}

{\pcomm} mentioned the hopping functions that are different from Gaussian form in {\mgb} and claimed that any such parametrization is uncontrolled. 
We note that the results for {\mgb} (that are consistent with the Gaussian approximation) have also been derived from the symmetry-representation method (for both the $\pi$-bonding states in App.\,H3a and the $\sigma$-bonding states in App.\,H5 of our work~\cite{Yu05032023GeometryEPC}), which does not assume any parametrization of the hopping function.
It means that the final results of different contributions to $\lambda$ hold regardless of the parametrization for the hopping functions, under our short-ranged hopping approximations (plus the small $k_x/k_y$ approximation for the $\sigma$-bonding states); the Gaussian approximation is a simple way to derive the results.
Moreover, we find that our EPC model can reasonably match the gauge invariant combinations of the EPC matrix elements along high-symmetry lines shown in Fig.\,9 of our work~\cite{Yu05032023GeometryEPC}, and the final $\lambda$ value that we get from our EPC model is reasonably close to the value directly from the {\abi} calculation.
Therefore, our theory is not uncontrolled.

\section{Having model parameters determined by {\abi} calculations is not an issue of a theory.}

{\pcomm} mentioned that several parameters in our model/expression (including the hopping parameters and $\langle \omega^2 \rangle$) are obtained from {\abi} calculations.
We note that it is not an issue for our electron or EPC model to have parameters determined by {\abi} calculations, similar to the fact that it is not an issue for an analytically constructed tight-binding model to have hopping parameters determined by {\abi} calculations.
For {\mgb}, we know $\sqrt{\langle \omega^2 \rangle}$ can be approximated by the frequency of $E_{2g}$ phonon at $\Gamma$ with about 10\% error.
Certainly, more understanding in the parameters can be gained by looking more into their expressions in the future, but those are not the focus of our current work.

\section{Summary}

In sum, the fundamental physics of {\mgb} has been included in our work, and the strong EPC strength for the $E_{2g}$ modes and $\sigma$-bonding states is manifested in the results derived from our EPC model.
Our results are not merely an outgrowth of the Gaussian approximation, and have meaningful implications on {\mgb} as well as future material search.

\end{document}